\documentclass[10pt, prl,aps,twocolumn,%
showpacs,preprintnumbers,amsmath,amssymb]{revtex4-2}
\usepackage[utf8]{inputenc}
\usepackage{graphicx}
\usepackage{bm}
\usepackage{physics}
\usepackage{siunitx}
\usepackage[breaklinks, colorlinks=true, citecolor=blue]{hyperref}

\newcommand{\Var}{\mathop{\mathrm{Var}}}
\newcommand{\Cov}{\mathop{\mathrm{Cov}}}

\begin{document}
\title{Proton Spin Fluctuation Decomposition}

\author{Tomoya Uji}
\email{uji@nt.phys.s.u-tokyo.ac.jp}
\affiliation{Department of Physics, The University of Tokyo, 7-3-1 Hongo, Bunkyo-ku, Tokyo 113-0033, Japan}

\begin{abstract}
    We formulate a fluctuation decomposition of proton spin in terms of the variances and covariances of the spin and orbital angular momentum of quarks and gluons.  We derive their representations in terms of single- and double-parton distributions.  The definite proton spin imposes sum rules requiring the variances of its spin and orbital contributions to be compensated by their covariances.  This framework reveals correlations and fluctuations that are invisible to the conventional proton spin decomposition.
\end{abstract}

\maketitle

\paragraph{Introduction:}
Understanding how quark and gluon angular momenta compose the proton spin is an important problem in quantum chromodynamics (QCD)~\cite{EuropeanMuon:1987isl}.  This question is commonly addressed through the proton spin decomposition, which expresses the total angular momentum as the sum of the spin and orbital angular momentum (OAM) of quarks and gluons~\cite{Jaffe:1989jz, Manohar:1990kr, Ji:1996ek, Bashinsky:1998if, Chen:2008ag, Wakamatsu:2010qj, Hatta:2011zs, Lorce:2011kd, Hatta:2011ku, Lorce:2012ce}. Each contribution is usually quantified by the expectation value of the corresponding operator in a polarized proton state.  However, the individual spin and OAM can fluctuate around their expectation values.

Such fluctuations reflect an elementary feature of quantum mechanics: an eigenstate of a sum of operators need not be an eigenstate of its individual terms.  Fluctuations and correlations reveal information beyond mean values in many areas of physics.  For example, charge and spin fluctuations in condensed-matter systems can probe entanglement between different spatial regions~\cite{Klich:2008un, Song:2010jgz, Song:2011gv}.  In mesoscopic transport, current correlations reveal the antibunching of electrons due to Fermi statistics~\cite{PhysRevB.46.12485, HBT:1999}.  In quantum optics, measurements of correlations in OAM and angular position reveal entanglement between photons~\cite{Mair:2001fyp, Leach:2010hqd}.  These examples illustrate how a fixed total quantity can coexist with fluctuating individual contributions, whose variances must be compensated by covariances.  For a proton with a definite spin projection, the same constraint applies to its partonic spin and OAM fluctuations.

Turning this observation into a QCD description requires information on correlations between partons.  Recent developments in double parton distributions (DPDs)~\cite{Diehl:2011tt, Diehl:2011yj, Manohar:2012jr} and their transverse-momentum-dependent and off-forward generalizations~\cite{Manohar:2012pe, Buffing:2017mqm, Panjsheeri:2024ysh} have advanced the theoretical description of such correlations: Spin correlations affect double-parton-scattering cross sections~\cite{Mekhfi:1985dv, Kasemets:2012pr, Echevarria:2015ufa, Cotogno:2018mfv}, progress in renormalization and evolution has clarified the scale dependence of DPDs~\cite{Gaunt:2009re, Manohar:2012pe, Diehl:2014vaa, Vladimirov:2016qkd, Buffing:2017mqm, Diehl:2018kgr}, and lattice calculations have begun to constrain DPD moments~\cite{Bali:2020mij, Bali:2021gel, Reitinger:2024ulw, Diehl:2025kcr}.  These developments provide concrete tools for studying variances and covariances of collective partonic quantities.

Studies of spin--orbit correlations (SOC) have already demonstrated the relevance of correlations beyond the average spin decomposition~\cite{Lorce:2011kd, Lorce:2014mxa, Kanazawa:2014nha, Mukherjee:2015aja, Rajan:2017cpx, Engelhardt:2021kdo, Kim:2024cbq, Hatta:2024otc, Bhattacharya:2024sck, Bhattacharya:2024wtg, Minamiguchi:2026usn}.  More recently, quantum-information approaches have explored entanglement among partons~\cite{Kovner:2015hga, Kharzeev:2017qzs, Beane:2019loz, Dumitru:2023fih, Qian:2024fqf, Dumitru:2025bib, Qian:2026aoh} and between spin and orbital degrees of freedom~\cite{Bhattacharya:2024sno, Hatta:2024lbw, Agrawal:2025yoe}.  A unified treatment of spin and orbital fluctuations also encompasses orbital variances and orbital--orbital covariances between different parton species, giving access to a broader set of physical questions.  Does a small average OAM imply little orbital motion?  Do the OAM of $u$ quarks and $\bar{u}$ antiquarks tend to align or anti-align?  These questions motivate organizing angular-momentum correlations according to the constraints imposed by the definite proton spin.  Here we formulate a fluctuation decomposition of proton spin, connecting the variances and covariances of its partonic contributions to single- and double-parton distributions.

\paragraph{Spin fluctuation decomposition:}
We consider a proton with four-momentum $P^\mu$ moving along the positive $z$-axis.  The proton state, $\ket{P, +}$, is an eigenstate of the total angular-momentum operator, $\hat{J}^z$, satisfying $\hat{J}^z\ket{P, +} = \frac{1}{2}\ket{P, +}$.  The operator, $\hat{J}^z$, is decomposed into the spin operator, $\hat{S}_a^z$, and the OAM operator, $\hat{L}_a^z$, of each parton species, $a$, as
\begin{equation}
    \hat{J}^z = \sum_a\left(\hat{S}_a^z + \hat{L}_a^z\right)\,, \qquad a\in\{q_f, \, \bar{q}_f, \, g\}\,,
    \label{eq:canonical_spin_decomposition}
\end{equation}
where $f$ labels quark flavors.  We adopt the gauge-invariant extension of the Jaffe--Manohar decomposition~\cite{Jaffe:1989jz, Manohar:1990kr, Bashinsky:1998if, Chen:2008ag, Hatta:2011zs, Lorce:2011kd, Hatta:2011ku, Lorce:2012ce}.  The spin and orbital operators are collectively denoted by $\hat{X}_A$, where the composite index $A$ specifies both the parton species and the type of angular momentum, with $\hat{X}_{(S, a)} = \hat{S}_a^z$ and $\hat{X}_{(L, a)} = \hat{L}_a^z$.

The fluctuation of each operator is $\delta\hat{X}_A = \hat{X}_A - \expval*{\hat{X}_A}$, where $\expval{\cdots}$ denotes an expectation value in the proton state.  The variance of each operator and the covariance between two operators are defined by
\begin{align}
    \Var(\hat{X}_A) &= \expval*{(\delta\hat{X}_A)^2}\,, \\
    \Cov(\hat{X}_A, \hat{X}_B) &= \frac{1}{2}\expval*{\{\delta\hat{X}_A, \delta\hat{X}_B\}}\,,
    \label{eq:angular_momentum_covariance}
\end{align}
where $\{\hat{X}_A, \hat{X}_B\} = \hat{X}_A\hat{X}_B + \hat{X}_B\hat{X}_A$ is the anticommutator.

The variances and covariances can distinguish different internal spin and orbital structures that give the same expectation values in the proton spin decomposition.  For example, if $\expval*{\hat{L}_{q_f}^z}$ is small, the net quark OAM may be narrowly distributed near zero or broadly distributed over positive and negative values that nearly cancel in the mean.  The variance $\Var(\hat{L}_{q_f}^z)$ distinguishes these possibilities.  Also, if $\Cov(\hat{S}_{q_f}^z, \hat{L}_g^z)$ is negative, an above-average quark spin tends to be accompanied by a below-average gluon OAM, indicating compensation between their fluctuations.

Since the proton has definite $J^z$, the fluctuation decomposition is
\begin{equation}
\begin{split}
    0 = \Var(\hat{J}^z) &= \sum_A\Var(\hat{X}_A) + \sum_{A\neq B}\Cov(\hat{X}_A, \hat{X}_B) \\
    &= \sum_{A, B}\Cov(\hat{X}_A, \hat{X}_B)\,.
    \label{eq:spin_fluctuation_decomposition}
\end{split}
\end{equation}
More generally, each operator, $\hat{X}_A$, has vanishing covariance with $\hat{J}^z$,
\begin{equation}
    \Cov(\hat{X}_A, \hat{J}^z) = \sum_B\Cov(\hat{X}_A, \hat{X}_B) = 0\,.
    \label{eq:component_covariance_sum_rule}
\end{equation}

\paragraph{Partonic representation:}
To give a partonic interpretation to the fluctuation decomposition, we resolve the spin and orbital contributions according to the longitudinal momentum fraction $x$ carried by a parton.  Quarks and antiquarks are treated as distinct species with positive momentum fractions.  In each light-front Fock component, the spin and orbital densities act as
\begin{align}
    \hat{S}_a^z(x) &= s_a\sum_{i\in a}\lambda_i\delta(x - x_i)\,, \\
    \hat{L}_a^z(x) &= \sum_{i\in a}\delta(x - x_i)\hat{\ell}_i\,,
    \label{eq:partonic_operator_densities}
\end{align}
where the sums run over all partons of species $a$, $x_i$ is the longitudinal momentum fraction of parton $i$, and $\lambda_i = \pm1$ is its helicity sign.  The spin weights are $s_{q_f} = s_{\bar{q}_f} = 1/2$ and $s_g = 1$.  The OAM of parton $i$ is $\hat{\ell}_i = (\hat{\bm{b}}_{i\perp}\times\hat{\bm{k}}_{i\perp})^z$.  Here $\hat{\bm{b}}_{i\perp}$ is the transverse position relative to the proton's transverse center of momentum, and $\hat{\bm{k}}_{i\perp}$ is the intrinsic transverse momentum~\cite{Burkardt:2002hr, Lorce:2011ni}.  Integrating $\hat{S}_a^z(x)$ and $\hat{L}_a^z(x)$ over $0 \leq x \leq 1$ recovers $\hat{S}_a^z$ and $\hat{L}_a^z$.

Taking proton expectation values of $\hat{S}_a^z(x)$ and $\hat{L}_a^z(x)$ connects the mean spin contribution to the helicity PDF, $\Delta f_a$~\cite{Manohar:1990kr, Bashinsky:1998if, Hatta:2011zs}, and the mean orbital contribution to a $\bm{k}_\perp$-moment of the GTMD, $F_{1,4}^a$~\cite{Lorce:2011kd, Hatta:2011ku},
\begin{align}
    \expval*{\hat{S}_a^z(x)} &= s_a\Delta f_a(x)\,,
    \label{eq:mean_spin_density} \\
    \expval*{\hat{L}_a^z(x)} &= - \int d^2\bm{k}_\perp\,\frac{\bm{k}_\perp^2}{M^2}F_{1,4}^a(x, 0, \bm{k}_\perp^2, 0, 0)\,,
    \label{eq:mean_orbital_density}
\end{align}
where $M$ is the proton mass.  The GTMD is evaluated at zero momentum transfer with a staple-shaped Wilson line~\cite{Lorce:2012ce}.

Resolving the parton momentum fractions allows us to identify which momentum regions are correlated with one another.  At fixed momentum fraction, the deviations from the mean are $\delta\hat{S}_a^z(x) = \hat{S}_a^z(x) - \expval*{\hat{S}_a^z(x)}$ and $\delta\hat{L}_a^z(x) = \hat{L}_a^z(x) - \expval*{\hat{L}_a^z(x)}$.  Correlations between the fluctuations at momentum fractions $x_1$ and $x_2$ are described by the covariance distributions
\begin{align}
    C_{SS}^{ab}(x_1, x_2) &= \frac{1}{2}\expval*{\{\delta\hat{S}_a^z(x_1), \delta\hat{S}_b^z(x_2)\}}\,,
    \label{eq:spin_spin_covariance_distribution}\\
    C_{SL}^{ab}(x_1, x_2) &= \frac{1}{2}\expval*{\{\delta\hat{S}_a^z(x_1), \delta\hat{L}_b^z(x_2)\}}\,,
    \label{eq:spin_orbit_covariance_distribution}\\
    C_{LL}^{ab}(x_1, x_2) &= \frac{1}{2}\expval*{\{\delta\hat{L}_a^z(x_1), \delta\hat{L}_b^z(x_2)\}}\,,
    \label{eq:orbital_orbital_covariance_distribution}
\end{align}
with $C_{LS}^{ab}(x_1, x_2) = C_{SL}^{ba}(x_2, x_1)$.  Integrating $C_{SS}^{ab}$, $C_{SL}^{ab}$, and $C_{LL}^{ab}$ over both momentum fractions recovers the component covariances entering Eq.~\eqref{eq:spin_fluctuation_decomposition}.  

To connect angular-momentum fluctuations with scattering observables, we express the spin--spin covariance, $C_{SS}^{ab}(x_1, x_2)$, in terms of PDFs and DPDs.  The two spin operators in $\hat{S}_a^z(x_1)\hat{S}_b^z(x_2)$ can act on the same parton or distinct partons.  The same-parton term involves the unpolarized PDF, $f_a(x)$.  The helicity correlation between distinct partons is described by
\begin{equation}
    D_{\Delta a\Delta b}(x_1, x_2) = \expval{\sum_{i\in a}\sum_{\substack{j\in b \\ j\neq i}}\lambda_i\lambda_j\delta(x_1 - x_i)\delta(x_2 - x_j)}\,.
    \label{eq:helicity_double_pdf}
\end{equation}
This distribution is related to the helicity DPD, $F_{\Delta a\Delta b}$~\cite{Diehl:2011yj, Diehl:2013mla}, by
\begin{equation}
    D_{\Delta a\Delta b}(x_1, x_2) = \int d^2\bm{y}_\perp\,F_{\Delta a\Delta b}(x_1, x_2, \bm{y}_\perp)\,,
    \label{eq:integrated_helicity_dpd}
\end{equation}
where $\bm{y}_\perp$ is the transverse separation between the two partons.

Combining above relation gives
\begin{equation}
\begin{split}
    C_{SS}^{ab}(x_1, x_2) &= \delta_{ab}s_a^2\delta(x_1 - x_2)f_a(x_1) \\
    &\qquad + s_as_bD_{\Delta a\Delta b}(x_1, x_2) \\
    &\qquad - s_as_b\Delta f_a(x_1)\Delta f_b(x_2)\,.
    \label{eq:spin_spin_partonic_representation}
\end{split}
\end{equation}
This identifies the parton distributions needed to determine the spin--spin covariance, providing a connection between spin fluctuations and experimentally accessible two-parton correlations.

We next consider the spin--orbit covariance, $C_{SL}^{ab}$.  Its same-parton term involves the correlation between a parton's helicity and its own OAM, described by
\begin{equation}
    P_a(x) = \expval{\sum_{i\in a}\lambda_i\delta(x - x_i)\hat{\ell}_i}\,.
    \label{eq:single_parton_spin_orbit_density}
\end{equation}
This is the canonical SOC~\cite{Lorce:2011kd, Lorce:2014mxa, Bhattacharya:2024sck}.  It is related to the single-parton GTMD, $G_{1,1}^a$, by
\begin{equation}
    P_a(x) = \int d^2\bm{k}_\perp\,\frac{\bm{k}_\perp^2}{M^2}G_{1,1}^a(x, 0, \bm{k}_\perp^2, 0, 0)\,.
    \label{eq:single_parton_soc_gtmd}
\end{equation}
The correlation between the helicity of one parton and the OAM of another is described by
\begin{equation}
    R_{ab}(x_1, x_2) = \expval{\sum_{i\in a}\sum_{\substack{j\in b \\ j\neq i}}\lambda_i\delta(x_1 - x_i)\delta(x_2 - x_j)\hat{\ell}_j}\,.
    \label{eq:distinct_parton_spin_orbit_distribution}
\end{equation}

To express $R_{ab}$ in terms of parton distributions, we retain the transverse momenta, $\bm{k}_{1\perp}$ and $\bm{k}_{2\perp}$, of the two active partons and their independent transverse momentum transfers, $\bm{\Delta}_{1\perp}$ and $\bm{\Delta}_{2\perp}$.  The corresponding double GTMD correlator, $W_{ab}^{rs}$~\cite{Diehl:2011yj}, is defined at zero skewness with the staple-link prescription.  The indices $r, s\in\{U, L\}$ specify the polarization projections of the first and second partons, respectively, with $U$ denoting unpolarized and $L$ longitudinally polarized partons.  Its explicit definition is given in the End Matter.

In $R_{ab}$, the first parton is weighted by its helicity and the second by its OAM.  The helicity weight is supplied by the longitudinal polarization projection in $W_{ab}^{LU}$.  The orbital weight, $(\bm{b}_{2\perp}\times\bm{k}_{2\perp})^z$, is generated by differentiating with respect to $\bm{\Delta}_{2\perp}$, which is Fourier conjugate to the second parton's position, $\bm{b}_{2\perp}$.  This gives
\begin{equation}
\begin{split}
    R_{ab}(x_1, x_2) &= -i\epsilon^{ij}\int d^2\bm{k}_{1\perp}\,d^2\bm{k}_{2\perp} \\
    &\qquad\qquad \times k_{2\perp}^j\frac{\partial W_{ab}^{LU}}{\partial\Delta_{2\perp}^i}\eval_{\bm{\Delta}_{1\perp} = \bm{\Delta}_{2\perp} = 0}\,,
\end{split}
    \label{eq:distinct_parton_spin_orbit_gtmd}
\end{equation}
where $\epsilon^{12} = +1$.  For the Wilson lines used here, the integration over $\bm{k}_{1\perp}$ reduces the first staple to a finite straight Wilson line, while the OAM projection on the second parton retains the staple-link prescription.  Parity and time-reversal symmetry imply that the projected distribution $R_{ab}$ is identical for future- and past-pointing staples, as for canonical OAM~\cite{Hatta:2011ku, Lorce:2012ce}.

Returning to the spin--orbit covariance, we obtain
\begin{equation}
\begin{split}
    C_{SL}^{ab}(x_1, x_2) &= \delta_{ab}s_a\delta(x_1 - x_2)P_a(x_1) \\
    &\qquad + s_aR_{ab}(x_1, x_2) \\
    &\qquad - s_a\Delta f_a(x_1)\expval*{\hat{L}_b^z(x_2)}\,.
\end{split}
    \label{eq:spin_orbit_partonic_representation}
\end{equation}

We finally consider the orbital--orbital covariance, $C_{LL}^{ab}$.  When both OAM operators act on the same parton, their product involves the squared OAM, $\hat{\ell}_i^2$.  The corresponding single-parton density is
\begin{equation}
    Q_a(x) = \expval{\sum_{i\in a}\delta(x - x_i)\hat{\ell}_i^2}\,.
    \label{eq:single_parton_orbital_square_density}
\end{equation}
Its representation in terms of the GTMD, $F_{1,1}^a$, is
\begin{align}
    Q_a(x) &= -2\int d^2\bm{k}_\perp\,\bm{k}_\perp^2\frac{\partial F_{1,1}^a(x, 0, \bm{k}_\perp^2, 0, \bm{\Delta}_\perp^2)}{\partial\bm{\Delta}_\perp^2}\eval_{\bm{\Delta}_\perp^2 = 0} \nonumber \\
    &\qquad - \frac{(1 - x)^2}{2}f_a(x)\,.
    \label{eq:single_parton_orbital_square_gtmd}
\end{align}
The derivative term gives the phase-space average of the squared orbital weight, $[(\bm{b}_\perp\times\bm{k}_\perp)^z]^2$, which corresponds to Weyl ordering of the transverse position and momentum operators~\cite{Belitsky:2003nz}.  The subtraction proportional to $f_a(x)$ converts this Weyl-ordered product into the operator square, $\hat{\ell}_i^2$, entering Eq.~\eqref{eq:single_parton_orbital_square_density}.  Its coefficient follows from the commutators, $[\hat{b}_{i\perp}^{m}, \hat{k}_{j\perp}^{n}] = i(\delta_{ij} - x_j)\delta^{mn}$.  Such a correction is absent in the spin--orbit sector, which involves only one OAM operator.

The OAM correlation between two distinct partons is described by
\begin{equation}
    U_{ab}(x_1, x_2) = \frac{1}{2}\expval{\sum_{i\in a}\sum_{\substack{j\in b \\ j\neq i}}\delta(x_1 - x_i)\delta(x_2 - x_j)\{\hat{\ell}_i, \hat{\ell}_j\}}\,.
    \label{eq:distinct_parton_orbital_distribution}
\end{equation}
Both parton helicities are summed over, selecting the unpolarized double GTMD, $W_{ab}^{UU}$.  The OAM weight of each parton is generated by differentiating with respect to its own transverse momentum transfer, giving
\begin{equation}
\begin{split}
    U_{ab}(x_1, x_2) &= -\epsilon^{ij}\epsilon^{mn}\int d^2\bm{k}_{1\perp}\,d^2\bm{k}_{2\perp}\,k_{1\perp}^j k_{2\perp}^n \\
    &\quad\times\frac{\partial^2 W_{ab}^{UU}}{\partial\Delta_{1\perp}^i\partial\Delta_{2\perp}^m}\eval_{\bm{\Delta}_{1\perp} = \bm{\Delta}_{2\perp} = 0} \\
    &\quad - \frac{x_1x_2}{2}D_{ab}(x_1, x_2)\,,
\end{split}
\label{eq:distinct_parton_orbital_gtmd}
\end{equation}
where the unpolarized DPD is obtained from $W_{ab}^{UU}$ as
\begin{equation}
    D_{ab}(x_1, x_2) = \int d^2\bm{k}_{1\perp}\,d^2\bm{k}_{2\perp}\,W_{ab}^{UU}\eval_{\bm{\Delta}_{1\perp} = \bm{\Delta}_{2\perp} = 0}\,.
    \label{eq:unpolarized_double_gtmd_reduction}
\end{equation}
The subtraction proportional to $D_{ab}$ is the same operator-ordering correction as the $f_a$-term in Eq.~\eqref{eq:single_parton_orbital_square_gtmd}.  Both OAM operators retain the staple-link prescription.  Parity and time-reversal symmetry imply invariance under simultaneous reversal of the two staples, but do not relate configurations with equal and opposite staple orientations.  When considering covariances, it is natural to define the OAM operators with a common Wilson-line prescription.  We therefore choose both staples to point to the same light-cone infinity.  The common direction is either future- or past-pointing without changing $U_{ab}$.

The orbital--orbital covariance is therefore expressed as
\begin{equation}
\begin{split}
    C_{LL}^{ab}(x_1, x_2) &= \delta_{ab}\delta(x_1 - x_2)Q_a(x_1) + U_{ab}(x_1, x_2) \\
    &\qquad - \expval*{\hat{L}_a^z(x_1)}\expval*{\hat{L}_b^z(x_2)}\,.
\end{split}
    \label{eq:orbital_orbital_partonic_representation}
\end{equation}
Together with the spin--spin and spin--orbit results, this completes the partonic representation of the fluctuation decomposition.

\paragraph{Sum rules:}
The sum rule in Eq.~\eqref{eq:component_covariance_sum_rule} constrains correlations involving a parton of species $a$ at momentum fraction $x_1$.  Substituting Eqs.~\eqref{eq:spin_spin_partonic_representation}, \eqref{eq:spin_orbit_partonic_representation}, and \eqref{eq:orbital_orbital_partonic_representation} into Eq.~\eqref{eq:component_covariance_sum_rule} gives the helicity sum rule
\begin{align}
    &s_af_a(x_1) + P_a(x_1) \nonumber \\
    &\quad + \sum_b\int_0^{1-x_1}dx_2\,\left[s_bD_{\Delta a\Delta b}(x_1, x_2) + R_{ab}(x_1, x_2)\right] \nonumber \\
    &\qquad = \frac{1}{2}\Delta f_a(x_1)\,.
    \label{eq:partonic_spin_sum_rule}
\end{align}
Each term on the left-hand side correlates the helicity of the selected parton with a different part of the total angular momentum.  The term $s_af_a$ is its correlation with its own spin.  The term $P_a$ correlates its helicity with its own OAM.  The distribution $s_bD_{\Delta a\Delta b}$ describes the correlation with the spin of another parton, while $R_{ab}$ describes the correlation with the OAM of another parton.  Summing these terms gives the correlation with the proton's total angular momentum, yielding $\frac{1}{2}\Delta f_a$ on the right-hand side.

Weighting the correlations by the selected parton's OAM instead of its helicity gives the analogous sum rule,
\begin{equation}
\begin{split}
    &s_aP_a(x_1) + Q_a(x_1) \\
    &\quad + \sum_b\int_0^{1-x_1}dx_2\,\left[s_bR_{ba}(x_2, x_1) + U_{ab}(x_1, x_2)\right] \\
    &\qquad = \frac{1}{2}\expval*{\hat{L}_a^z(x_1)}\,,
\end{split}
    \label{eq:partonic_orbital_sum_rule}
\end{equation}
which constrains the compensation of its OAM fluctuations in the same manner.


\paragraph{Illustrations and constraints:}
To illustrate the fluctuation decomposition in a perturbatively calculable system, we replace the proton by a quark of mass $m$ dressed by gluon radiation~\cite{Brodsky:2000ii, Kanazawa:2014nha, Mukherjee:2014nya, Mukherjee:2015aja, Kasemets:2016nio}.  We choose the state with $J^z = 1/2$, retain its one-quark and quark--gluon light-front Fock components, and evaluate the variances and covariances through $O(\alpha_s)$.  This provides a simple setting in which the spin and OAM of quarks and gluons fluctuate while their sum remains fixed.

In the two-parton sector, $x$ is the quark momentum fraction and $1 - x$ is the gluon momentum fraction.  The probability density for finding a quark--gluon pair with helicity signs $(\lambda_q, \lambda_g)$ and quark momentum fraction $x$ in the dressed-quark state is denoted by $p_{\lambda_q\lambda_g}(x)$.  The three nonzero densities are $p_{++}$, $p_{+-}$, and $p_{-+}$, while $p_{--} = 0$.  These three probability densities determine all single- and two-parton distributions entering the fluctuation decomposition.  Explicit expressions for the probabilities and distributions are given in the End Matter.

\begin{figure}
    \centering
    \includegraphics[width=\columnwidth]{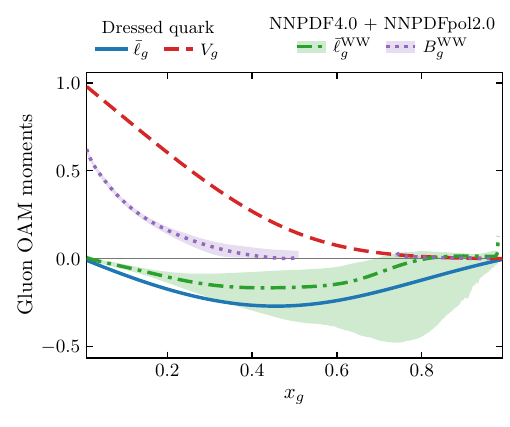}
    \caption{Conditional gluon OAM mean $\bar{\ell}_g$ (blue solid) and variance $V_g$ (red dashed) in the dressed-quark model with $\Lambda_{\mathrm{UV}}/m=2$ and $\varepsilon=0.01$.  Also shown are the mean $\bar{\ell}_g^{\mathrm{WW}}$ (green dash-dotted) and the lower bound on the variance $B_g^{\mathrm{WW}}$ (purple dotted) for the proton, using NNPDF4.0 and NNPDFpol2.0.  Shading indicates pointwise 68\% PDF uncertainty bands; see the End Matter for details.}
    \label{fig:gluon_oam_variance}
\end{figure}

Using these distributions, we examine two examples of the information contained in angular-momentum fluctuations.  We first consider OAM for a gluon with longitudinal momentum fraction $x_g = 1 - x$.  Its conditional mean, $\bar{\ell}_g(x_g)$, and variance, $V_g(x_g)$, are
\begin{align}
    \bar{\ell}_g(x_g) &= \frac{\expval*{\hat{L}_g^z(x_g)}}{f_g(x_g)}\,,
    \label{eq:dq_conditional_gluon_oam_mean} \\
    V_g(x_g) &= \frac{Q_g(x_g)}{f_g(x_g)} - \overline{\ell}_g^2(x_g)\,.
    \label{eq:dq_conditional_gluon_oam_variance}
\end{align}
With the normalization by $f_g(x_g)$, $\bar{\ell}_g(x_g)$ gives the average OAM of a gluon found with momentum fraction $x_g$, while $V_g(x_g)$ measures the fluctuations of its OAM about that average.

Figure~\ref{fig:gluon_oam_variance} shows the conditional mean and variance in the dressed-quark state.  At small $x_g$, $p_{+-}\simeq p_{++}$ and $p_{-+}$ is relatively suppressed, so the gluon carries OAM close to $-1$ or $+1$ with nearly equal probabilities.  These opposite orbital motions nearly cancel in the mean, giving $\bar{\ell}_g(x_g)\to0$, while $V_g(x_g)\to1$.  As $x_g\to1$, all allowed gluon OAM values, $0$ and $\pm(1-x_g)$, approach zero, so both the conditional mean and variance vanish.  The variance thus distinguishes two different origins of a small mean OAM: cancellation between opposite orbital motions and an OAM distribution concentrated near zero.

For comparison, Fig.~\ref{fig:gluon_oam_variance} also includes estimates based on the unpolarized gluon PDF from NNPDF4.0~\cite{NNPDF:2021njg} and the helicity PDF from NNPDFpol2.0~\cite{Cruz-Martinez:2025ahf}.  The conditional mean, $\bar{\ell}_g^{\mathrm{WW}}$, is evaluated in the Wandzura--Wilczek (WW) approximation.  The variance, $V_g$, requires the single-parton squared-OAM density $Q_g$ in Eq.~\eqref{eq:single_parton_orbital_square_density}, which these PDF inputs do not determine.  The helicity PDF and SOC nevertheless provide a lower bound: applying the Cauchy--Schwarz inequality to the helicity and OAM fluctuations in the conditional partonic state gives
\begin{equation}
    V_g(x_g)\geq B_g(x_g)=\frac{\left[P_g(x_g)-\Delta f_g(x_g)\bar{\ell}_g(x_g)\right]^2}{f_g^2(x_g)-[\Delta f_g(x_g)]^2}\,,
    \label{eq:conditional_gluon_oam_bound}
\end{equation}
for $f_g(x_g)>0$ and $|\Delta f_g(x_g)|<f_g(x_g)$.

Using the WW expressions for the OAM and SOC densities, we estimate this lower bound as $B_g^{\mathrm{WW}}$, shown in Fig.~\ref{fig:gluon_oam_variance}.  The central $B_g^{\mathrm{WW}}$ curve is omitted where $|\Delta g/g|$ approaches or exceeds unity, making the helicity-variance denominator small or nonpositive.  The integral relations, PDF inputs, and uncertainty prescription are detailed in the End Matter.  At small $x_g$ in the plotted range, the estimated bound remains appreciable while $\bar{\ell}_g^{\mathrm{WW}}$ is close to zero, indicating substantial orbital fluctuations within this approximation.  The decrease of $B_g^{\mathrm{WW}}$ at larger $x_g$ weakens the constraint.

We next examine how quark OAM fluctuations are compensated by correlations with the other angular-momentum components.  For $x = 1 - x_g$, the relevant covariances are
\begin{equation}
\begin{aligned}
    \Cov(\hat{L}_{q_f}^z(x), \hat{L}_{q_f}^z) &= Q_{q_f}(x)\,, \\
    \Cov(\hat{L}_{q_f}^z(x), \hat{L}_g^z) &= \int_{\varepsilon}^{1-\varepsilon}dx_g'\,U_{q_fg}(x, x_g')\,, \\
    \Cov(\hat{L}_{q_f}^z(x), \hat{S}_g^z) &= \int_{\varepsilon}^{1-\varepsilon}dx_g'\,R_{gq_f}(x_g', x)\,.
\end{aligned}
    \label{eq:dq_quark_oam_covariances}
\end{equation}
Since $P_{q_f}(x) = \expval*{\hat{L}_{q_f}^z(x)}$, the orbital sum rule, Eq.~\eqref{eq:partonic_orbital_sum_rule}, requires the three covariances to sum to zero.

\begin{figure}
    \centering
    \includegraphics[width=\columnwidth]{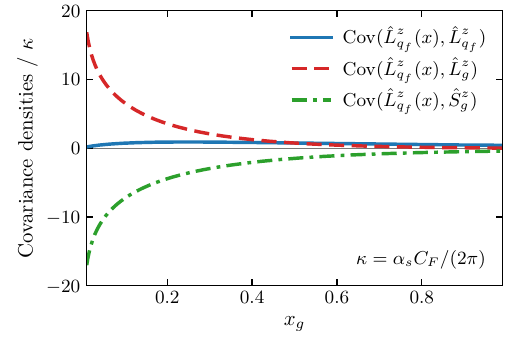}
    \caption{Covariances of the quark OAM at $x = 1 - x_g$ with the integrated quark OAM (blue solid), gluon OAM (red dashed), and gluon spin (green dash-dotted), given in Eq.~\eqref{eq:dq_quark_oam_covariances}.  Results are shown for $\Lambda_{\mathrm{UV}}/m = 2$ and longitudinal cutoff $\varepsilon = 0.01$, with all covariances divided by $\kappa = \alpha_s C_F/(2\pi)$.}
    \label{fig:quark_oam_covariance}
\end{figure}

Figure~\ref{fig:quark_oam_covariance} shows a near cancellation between the positive covariance of quark OAM with gluon OAM and the negative covariance with gluon spin at small $x_g$.  The soft gluon carries almost all of the relative OAM, opposite to its spin, consistent with the small-$x$ gluon SOC, $P_g(x_g) / f_g(x_g) \to -1$~\cite{Bhattacharya:2024sno}.  In this two-parton state, the shared relative orbital motion relates the known soft-gluon spin--OAM anti-alignment to correlations with the recoiling quark.  The fluctuation decomposition makes explicit how the sum of these correlations compensates the quark OAM variance density.

The detailed dressed-quark results presented here depend on the model.  These examples nevertheless demonstrate that the fluctuation decomposition reveals a correlated structure of partonic angular momentum that is invisible to the conventional spin decomposition.

\paragraph{Conclusion:}
We have formulated a fluctuation decomposition of proton spin and expressed its variances and covariances in terms of single- and double-parton distributions.  The resulting sum rules constrain how fluctuations of individual spin and orbital contributions are compensated by correlations with the remaining components.

The sum rules can also be used to constrain combinations of double-GTMD moments from independent information on single-parton distributions and DPDs.  An important next step is to identify scattering observables sensitive to these moments and to establish the corresponding factorization.  Another direction is to explore whether these moments admit a representation in terms of higher-twist double GPDs~\cite{Manohar:2012pe} and related multiparton correlators, extending the known relations between canonical OAM and twist-three GPDs~\cite{Hatta:2012cs, Rajan:2017cpx, Hatta:2024otc}.  Such relations could provide a complementary approach to nonperturbative calculations and the investigation of experimental access to orbital covariances.  Determining these correlations in the proton would reveal how fluctuating partonic spin and OAM combine to give a definite total angular momentum.

\hspace{3pt}
\begin{acknowledgments}
The author thanks Yuki~Fujimoto, Kenji~Fukushima, and Josuke~Minamiguchi for useful discussions.  This work was partially supported by FoPM, WINGS Program, the University of Tokyo.
\end{acknowledgments}

\bibliography{reference}

\clearpage

\section*{End Matter}
\label{sec:end_matter}

\paragraph{Double GTMD correlator:}
The double GTMD correlator introduced in Eqs.~\eqref{eq:distinct_parton_spin_orbit_gtmd} and \eqref{eq:distinct_parton_orbital_gtmd} is defined by~\cite{Diehl:2011yj}
\begin{equation}
\begin{split}
    &W_{ab}^{rs}(x_1, \bm{k}_{1\perp}; x_2, \bm{k}_{2\perp}; \bm{\Delta}_{1\perp}, \bm{\Delta}_{2\perp}) \\
    =& \frac{2P^+}{(x_1P^+)^{n_a}(x_2P^+)^{n_b}}\int dy^-d^2\bm{y}_\perp\,e^{i(\bm{\Delta}_{1\perp} - \bm{\Delta}_{2\perp})\cdot\bm{y}_\perp/2} \\
    &\qquad\times\prod_{m=1}^{2}\left[\int\frac{dz_m^-\,d^2\bm{z}_{m\perp}}{(2\pi)^3}\,e^{ix_mP^+z_m^- - i\bm{k}_{m\perp}\cdot\bm{z}_{m\perp}}\right] \\
    &\qquad\times\mel**{p', +}{\left[\hat{\mathcal{O}}_b^s\left(-\frac{y}{2}, z_2\right)\hat{\mathcal{O}}_a^r\left(\frac{y}{2}, z_1\right)\right]}{p, +}\,,
\end{split}
    \label{eq:double_gtmd_correlator}
\end{equation}
where $y^+ = z_1^+ = z_2^+ = 0$, $n_{q_f} = n_{\bar{q}_f} = 0$, and $n_g = 1$.  The indices $r, s\in\{U, L\}$ specify the polarization projections of the first and second partons, respectively, with $U$ denoting unpolarized and $L$ longitudinally polarized partons.  The operators $\hat{\mathcal{O}}_a^{U, L}$ are color-singlet bilocals with the same staple-link prescription.  We work at zero skewness and choose the symmetric transverse frame, so that $p^{\prime +} = p^+ = P^+$, $\bm{p}_\perp = -\bm{\Delta}_\perp/2$, and $\bm{p}'_\perp = \bm{\Delta}_\perp/2$, with $\bm{\Delta}_\perp = \bm{\Delta}_{1\perp} + \bm{\Delta}_{2\perp}$.  The normalization is fixed so that the forward transverse-momentum integrals of $W_{ab}^{UU}$ and $W_{ab}^{LL}$ give $D_{ab}$ and $D_{\Delta a\Delta b}$, respectively.

\paragraph{Dressed-quark model:}

\begin{table}
    \caption{Single-parton distributions in the dressed-quark state.  All probabilities are evaluated at the quark momentum fraction $x$.}
    \label{tab:dq_single_distributions}
    \begin{ruledtabular}
    \begin{tabular}{ll}
        Quantity & Result \\
        \hline
        $f_{q_f}(x)$ & $Z\delta(1-x) + p_{++} + p_{+-} + p_{-+}$ \\
        $\Delta f_{q_f}(x)$ & $Z\delta(1-x) + p_{++} + p_{+-} - p_{-+}$ \\
        $f_g(1-x)$ & $p_{++} + p_{+-} + p_{-+}$ \\
        $\Delta f_g(1-x)$ & $p_{++} - p_{+-} + p_{-+}$ \\
        $\expval*{\hat{L}_{q_f}^z(x)} = P_{q_f}(x)$ & $(1-x)(p_{+-} - p_{++})$ \\
        $\expval*{\hat{L}_g^z(1-x)}$ & $x(p_{+-} - p_{++})$ \\
        $P_g(1-x)$ & $-x(p_{++} + p_{+-})$ \\
        $Q_{q_f}(x)$ & $(1-x)^2(p_{++} + p_{+-})$ \\
        $Q_g(1-x)$ & $x^2(p_{++} + p_{+-})$
    \end{tabular}
    \end{ruledtabular}
\end{table}

\begin{table}
    \caption{Two-parton distributions in the dressed-quark state.  All probabilities are evaluated at the quark momentum fraction $x_1$.}
    \label{tab:dq_pair_distributions}
    \begin{ruledtabular}
    \begin{tabular}{ll}
        Quantity & Result \\
        \hline
        $D_{q_fg}(x_1, x_2)$ & $(p_{++} + p_{+-} + p_{-+})\delta(1 - x_1 - x_2)$ \\
        $D_{\Delta q_f\Delta g}(x_1, x_2)$ & $(p_{++} - p_{+-} - p_{-+})\delta(1 - x_1 - x_2)$ \\
        $R_{q_fg}(x_1, x_2)$ & $x_1(p_{+-} - p_{++})\delta(1 - x_1 - x_2)$ \\
        $R_{gq_f}(x_2, x_1)$ & $-(1-x_1)(p_{++} + p_{+-})\delta(1 - x_1 - x_2)$ \\
        $U_{q_fg}(x_1, x_2)$ & $x_1(1-x_1)(p_{++} + p_{+-})\delta(1 - x_1 - x_2)$
    \end{tabular}
    \end{ruledtabular}
\end{table}

We use the light-front wave functions of Ref.~\cite{Kanazawa:2014nha} and regulate their transverse-momentum dependence as
\begin{align}
    \frac{1}{\bm{k}_\perp^2 + m^2(1 - x)^2}
    &\longrightarrow \frac{1}{\bm{k}_\perp^2 + m^2(1 - x)^2} \label{eq:dq_regulator} \\
    &\qquad - \frac{1}{\bm{k}_\perp^2 + m^2(1 - x)^2 + \Lambda_{\mathrm{UV}}^2}\,, \nonumber
\end{align}
where $m$ is the quark mass, and $\Lambda_{\mathrm{UV}}$ is the transverse regulator scale.  The regulator leaves the helicity labels and azimuthal dependence of the wave functions unchanged, so that the regulated state remains an eigenstate of $\hat{J}^z$ with eigenvalue $1/2$.  Integrating the squared regulated wave functions over transverse momentum gives
\begin{equation}
\begin{split}
    p_{++}(x) &= \frac{\alpha_s C_F}{2\pi}\frac{I_1(x)}{1 - x}\,, \\
    p_{+-}(x) &= x^2p_{++}(x)\,, \\
    p_{-+}(x) &= \frac{\alpha_s C_F}{2\pi}m^2(1 - x)^3I_0(x)\,,
    \label{eq:dq_probabilities}
\end{split}
\end{equation}
and $p_{--}(x) = 0$.  Here $C_F = (N_c^2 - 1)/(2N_c)$, and 
\begin{align}
    I_0(x) &= \frac{1}{m^2(1 - x)^2} + \frac{1}{m^2(1 - x)^2 + \Lambda_{\mathrm{UV}}^2} \nonumber \\
    &\qquad - \frac{2}{\Lambda_{\mathrm{UV}}^2}\ln\left[1 + \frac{\Lambda_{\mathrm{UV}}^2}{m^2(1 - x)^2}\right]\,, \label{eq:dq_radial_integrals} \\
    I_1(x) &= \left[1 + \frac{2m^2(1 - x)^2}{\Lambda_{\mathrm{UV}}^2}\right]\ln\left[1 + \frac{\Lambda_{\mathrm{UV}}^2}{m^2(1 - x)^2}\right] - 2\,. \nonumber
\end{align}
The two-parton probabilities are restricted to $\varepsilon < x < 1 - \varepsilon$ and vanish outside this interval, where $0 < \varepsilon < 1/2$ is a common longitudinal cutoff.  The one-quark probability is fixed by normalization,
\begin{equation}
\begin{split}
    Z &= 1 - \int_\varepsilon^{1-\varepsilon}dx\,
    \left[p_{++}(x) + p_{+-}(x) + p_{-+}(x)\right]\,,
\end{split}
    \label{eq:dq_normalization}
\end{equation}
and contributes $Z\delta(1 - x)$ to both quark PDFs.

The resulting single- and two-parton distributions are summarized in Tables~\ref{tab:dq_single_distributions} and \ref{tab:dq_pair_distributions}.  The two-parton distributions have support on $x_1 + x_2 = 1$, since the quark and gluon carry the entire longitudinal momentum of the two-parton state.  The quark--quark and gluon--gluon two-parton distributions vanish at this order, as do all distributions involving antiquarks or other quark flavors.  The unpolarized and helicity distributions, $D_{q_fg}$ and $D_{\Delta q_f\Delta g}$, agree with the corresponding results of Ref.~\cite{Kasemets:2016nio} when the same regulator is applied to the wave functions and the transverse separation is integrated out.

\paragraph{Proton inputs for Fig.~\ref{fig:gluon_oam_variance}:}
We write $g(x) = f_g(x)$, $\Delta g(x) = \Delta f_g(x)$, and $L_g(x) = \expval*{\hat{L}_g^z(x)}$.  The WW approximation retains the twist-two contributions to the canonical OAM and SOC densities and neglects genuine twist-three correlators~\cite{Hatta:2012cs, Bhattacharya:2024sck, Hatta:2024otc}.  With the positive-$x$ normalization used here, these densities are
\begin{align}
    L_g^{\mathrm{WW}}(x) &= x\int_x^1\frac{dz}{z^2}\left[H_g(z,0,0)+E_g(z,0,0)\right] \nonumber \\
    &\qquad -2x\int_x^1\frac{dz}{z^2}\,\Delta g(z)\,,
    \label{eq:ww_gluon_oam} \\
    P_g^{\mathrm{WW}}(x) &= x\int_x^1 dz\left[\frac{\Delta g(z)}{z}-\frac{2g(z)}{z^2}\right]\,.
    \label{eq:ww_gluon_soc}
\end{align}
Here $H_g(z,0,0) = zg(z)$.  For the numerical estimate we set $E_g(z,0,0)=0$ and use the leading-order sets \texttt{NNPDF40\_lo\_pch\_as\_01180}~\cite{NNPDF:2021njg} and \texttt{NNPDFpol20\_lo\_as\_01180}~\cite{Cruz-Martinez:2025ahf} at the common factorization scale $\mu^2=10\,\mathrm{GeV}^2$.

The plotted mean is $\bar{\ell}_g^{\mathrm{WW}}(x)=L_g^{\mathrm{WW}}(x)/g(x)$, and the corresponding estimate of the variance bound is
\begin{equation}
    B_g^{\mathrm{WW}}=\frac{\left[P_g^{\mathrm{WW}}-(\Delta g/g)L_g^{\mathrm{WW}}\right]^2}{g^2-(\Delta g)^2}\,,
    \label{eq:ww_gluon_oam_bound}
\end{equation}
where all functions are evaluated at the same $x$ and $\mu$.  The ratio is applicable only for $g>0$ and $|\Delta g|<g$.  Equation~\eqref{eq:conditional_gluon_oam_bound} is an inequality for consistently defined exact conditional moments, whereas Eq.~\eqref{eq:ww_gluon_oam_bound} estimates its right-hand side using WW inputs.

The central curves in Fig.~\ref{fig:gluon_oam_variance} use member 0 of each PDF set.  For the bands, we combine all $100\times100$ replica pairs with equal weight, treating the two ensembles as independent.  Each replica pair is used consistently throughout the integrals and ratios, preserving its correlations in $x$.  At each external $x$, the bands span the 16th--84th percentiles of retained finite values: $\bar{\ell}_g^{\mathrm{WW}}$ requires $g > 0$, while $B_g^{\mathrm{WW}}$ additionally requires $|\Delta g| < g$.

\end{document}